\documentclass[prc,twocolumn,floatfix,nofootinbib]{revtex4-1}

\usepackage{graphicx}
\usepackage{amssymb,amsmath,amstext,amsthm,amsfonts}
\usepackage{slashed}
\usepackage{epstopdf}

\usepackage[usenames]{color}
\definecolor{magenta}{cmyk}{0,1,0,0}

\definecolor{blue}{rgb}{0,0,1}

\newcommand{\onepion}{single-pion\ }

\begin{document}

\title{Pion production in the T2K experiment}
\author{O. Lalakulich}
\author{U. Mosel}
\email[Contact e-mail: ]{mosel@physik.uni-giessen.de}
\affiliation{Institut f\"ur Theoretische Physik, Universit\"at Giessen, D-35392 Giessen, Germany}

\begin{abstract}
\begin{description}
\item[Background] Pion production gives information on the axial form factors of
nucleon resonances. It also introduces a noticeable background to quasi-elastic
measurements on nuclear targets and thus has a significant impact on precision
studies of neutrino oscillation parameters.
\item[Purpose] To clarify neutrino-induced pion production on nucleons and nuclei.
\item[Method] The Giessen Boltzmann--Uehling--Uhlenbeck (GiBUU) model is used for the description of neutrino-nucleus reactions.
\item[Results] Theoretical results for differential cross sections for the T2K
neutrino flux at the ND280 detector and integrated cross sections as a function of neutrino energy are given. Two sets of pion production data on
elementary targets are used as inputs to obtain limits for pion production in neutrino-nucleus
reactions.
\item[Conclusions] Pion production in the T2K ND280 detector can help to narrow down the uncertainties in the
elementary pion production cross sections. It can also give valuable information
on the nucleon-$\Delta$ axial form factor.
\end{description}
\end{abstract}

\date{\today}

\maketitle

\section{Introduction}
Pion production in neutrino-nucleus reactions represents one of the main
backgrounds to the identification of charged current quasielastic (CC QE)
scattering; the latter is used as a tool to reconstruct the neutrino energy. Pions in the
final state that go unobserved for some experimental
reasons contribute to the background as well as pions produced in the initial reaction that are absorbed in the
nucleus (so-called stuck-pion events).
Experimentalists are well aware of this latter complication  and have
subtracted the cross sections for such events from their original data. This is
done with the help of event generators such as NUANCE (for MiniBooNE) or GENIE
or NEUT (for T2K) that are tuned to experimental pion data obtained
in the same experiment. The quality of the measured pion production cross
sections as well as the quality of the generators thus directly affect the final
QE data. Furthermore, the subtraction of stuck-pion events from the
QE-like ones involves the reconstruction of neutrino energies which itself can distort the
cross sections \cite{Lalakulich:2012hs}.
In \cite{Leitner:2010kp} it has been shown that even in the absence of 2p-2h or deep inelastic scattering (DIS) processes
the energy reconstruction is affected by pion
events. For example, for
Cerenkov type detectors the reconstructed energy exhibit a bump at values
lower than the true neutrino energy, this bump being dependent on the pion
detection threshold (see Figs.\ 6 and 7 in \cite{Leitner:2010kp}).

For experiments in the current era of precise measurements it is,
therefore, important to have the pion production well under control. In
\cite{Lalakulich:2012cj} we have
performed a detailed comparison of theoretical calculations, which use
state-of-the-art primary pion production models and final state interactions,
with the MiniBooNE data \cite{AguilarArevalo:2010bm,AguilarArevalo:2010xt}. The
calculated cross sections are consistently lower than the
experimental ones and show a different kinetic energy (or momentum)
distributions of pions; both of these features have recently been confirmed by
Hernandez et al. \cite{Hernandez:2013jka}.

The major source of uncertainty in our theoretical calculations is the
elementary pion production cross section used as input. Currently it is only known
with at least a 30\% level of uncertainty, which is based on the old
bubble-chamber data on proton and deuterium targets. Only the upper boundary of
this uncertainty range (BNL data \cite{Kitagaki:1986ct} as opposed to those
obtained at ANL \cite{Radecky:1981fn}) comes close to the MiniBooNE data.
A definite statement on the validity of the elementary cross sections is,
however, not possible since also flux uncertainties could be responsible for
the observed disagreement. In addition, also 2p2h1$\pi$ processes could play a
role \cite{Lalakulich:2012cj}. A first attempt in this direction has recently been
undertaken in Ref.\ \cite{Mariano:2013dwa}.

\begin{figure}
\includegraphics[angle=-90,width=0.5\textwidth]{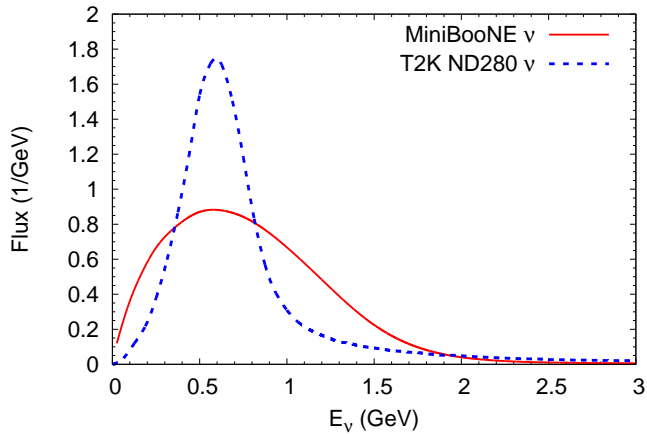}
\caption{(Color online) Flux distributions for the experiments MiniBooNE (solid
line) and T2K (dashed line), both normalized to 1 when integrated over their
full range. The long flat tail of the T2K flux ranging up to 22 GeV is not
shown. The distributions are taken from \cite{AguilarArevalo:2008yp} and
\cite{t2k}. } \label{fig:flux}
\end{figure}

In this situation any independent experiment could clearly help to clarify the
situation. Therefore in this brief report we give the results for pion
production in the T2K experiment, using the neutrino flux recently published in
Ref.~\cite{t2kflux}. While the peak energy of T2K is similar to that of
MiniBooNE (around 600 MeV) the T2K flux distribution is significantly
narrower, as shown in Fig.~\ref{fig:flux}. As a consequence, less influence of
RPA correlations is expected (because of the smaller weight of lower energies)
and, in addition, pion production is expected to be less prevalent (because of
the suppression of higher energies) in T2K.

All the results discussed in this paper are obtained within the GiBUU model
\cite{Buss:2011mx}. All technical details for pion production and an extended
discussion of in-medium effects can be found in Ref.\ \cite{Lalakulich:2012cj}
devoted to the comparison with the MiniBooNE data,
the only difference being the different neutrino flux distribution
used here, namely the T2K flux.
We stress that there is no tuning of parameters of any kind. The calculations
contain QE (with an axial mass of 1 GeV), 2p-2h processes determined with the
help of the MiniBooNE data, pion production through resonances and DIS.

For completeness we note here that the present calculations deal only with the
incoherent part of the pion production cross section. Coherent pion production
from nuclear targets requires phase coherence and can thus not be treated with a
transport (or Monte Carlo) description. Generators that contain such contributions use
oversimplified descriptions for the coherent pion production cross section (for
a discussion see \cite{Morfin:2009zz}). A calculation that is free of such
oversimplifications gives for a $^{12}$C target a cross section of about $0.03
\times 10^{-38}\mbox{cm}^2$ at the peak T2K energy of 0.6 GeV
\cite{Nakamura:2009iq}.

\section{Pion production cross sections}
The calculations give a flux-averaged total inclusive cross section of $8.32
\times 10^{-39} \mbox{cm}^2$ per nucleon at an average energy of 0.93 GeV.
This point fits very nicely into the systematics shown in Fig.\ 13 in Ref.\
\cite{Abe:2013jth}\footnote{The cross section calculated here is higher than the
experimental value of $6.91 \pm 0.13 \pm 0.84 \times 10^{-39} \mbox{cm}^2$ per
nucleon at the mean energy of 0.85 GeV \cite{Abe:2013jth}.
We note that we used the flux taken from \cite{t2kflux} and our value of the
mean neutrino energy is calculated over the full range of this flux, i.e. up to
22 GeV. This slightly differs from the flux
reported in \cite{Abe:2013jth,inclflux} and their energy range up to 10 GeV
only.}. Our calculated cross section is made up of true quasielastic (QE) scattering (3.68), 2p-2h (0.95),
$\Delta$ excitation (1.72), higher resonance excitation (0.25), \onepion
background (0.44) and DIS (1.29); the numbers in parentheses give the partial
cross sections per nucleon in units of $10^{-39} \mbox{cm}^2$. At first sight
the relatively large DIS
contribution is surprising since the T2K flux is peaked at only 0.6 GeV; it has,
however, a weak, but long tail all the way up to 22 GeV and the DIS cross
section goes linear with energy.
From the theoretical side,  the pion production cross section is  sensitive to the prescription
how to describe the transition from resonance-dominated to DIS-dominated physics
around invariant masses of about 2 GeV.
Here we have used the same prescription (see Ref.~\cite{Lalakulich:2012gm}) as in our
earlier calculations.

\begin{figure}
\includegraphics[width=0.45\textwidth]{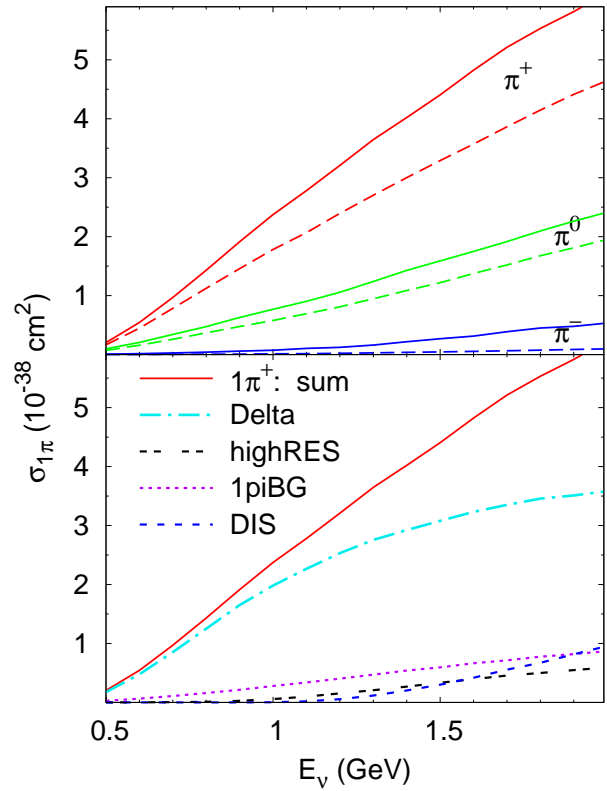}
\caption{(Color online) The cross section for \onepion production on $^{12}$C
as a function of
neutrino energy. In the upper part the cross sections are shown for all three
charge states for BNL \cite{Kitagaki:1986ct} (upper) and ANL \cite{Radecky:1981fn} (lower) inputs. The solid curves give
the results of a calculation with the BNL input, the dashed one that with the
ANL input. In the lower part of the figure the contributions from various
processes are shown for 1$\pi^+$ production. Here the BNL cross sections have
been used as elementary input.}
\label{fig:pionEdep}
\end{figure}

The calculated total pion-production cross sections
are: 0.65 for 1$\pi^-$ production, 1.53 for 1$\pi^0$  and 2.53
for 1$\pi^+$
(again all per nucleon, in units of $10^{-39} \mbox{cm}^2$), these numbers can
be compared with 5.0 for neutron and 11.2 for proton knock-out. The comparison
with the MiniBooNE data in \cite{Lalakulich:2012cj}  showed that the calculated
values, mainly for 1$\pi^+$, were in general somewhat too small compared to the
data. To facilitate the comparison we, therefore, give in Fig.\
\ref{fig:pionEdep} the integrated \onepion production cross section for $^{12}$C
target plotted as a function of true neutrino energy. We also show the
contributions of
the various reaction mechanisms in the lower part of that figure.
Fig.\ \ref{fig:pionEdep} shows that pion production through
the $\Delta$ resonance nearly exhausts the cross section up to a neutrino
energy of about 0.8 GeV, with various other components becoming significant
above that energy. DIS becomes dominant above about 2 GeV.

As we have discussed in \cite{Lalakulich:2012cj}, the neutrino energy
reconstruction works quite well for pion production when it is based on
muon-pion kinematics (assuming that 2p2h1$\pi$ processes play only a minor
role). If T2K could collect enough statistics to limit the
(reconstructed) energy up to about 0.8 GeV, this would give fairly clean
information on the $N\Delta$ coupling.
In this energy range the pion production
cross section is expected to be at the level of 0.2 - 1.5 $\times 10^{-38} \,
\mbox{cm}^2$ and to be dominated by the production and the following
decay of the $\Delta$ resonance (see Fig.\ \ref{fig:pionEdep}, lower part). Therefore,
the absolute value of the cross section will reveal information on the coupling,
even when the kinematics of the final pion is modified via final state
interactions in a carbon nucleus.

\begin{figure}
\includegraphics[angle=-90,width=0.5\textwidth]{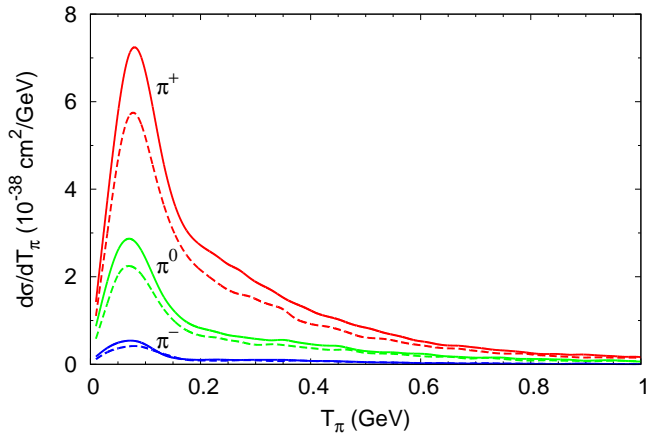}
\caption{(Color online) Kinetic energy spectra for pions of indicated charge
calculated for the T2K flux \cite{t2kflux}. In the pairs of curves the upper,
solid curve always gives the results obtained with the BNL input, the lower,
dashed one that obtained with the ANL input.} \label{fig:pion-spectra}
\end{figure}

In Fig.\ \ref{fig:pion-spectra} we show the pion kinetic energy spectra for all three pion charge states.
All the spectra show the well-known shape, with a peak at around $T_\pi = 80$
MeV extending up to about 200 MeV to be followed by a broad shoulder towards
higher values of $T_\pi$. This shape has also been observed in
pion-photoproduction
experiments on nuclei \cite{Krusche:2004uw}. The steep falloff at the right-hand
shoulder of the peak is a consequence of pion reabsorption in carbon, which
happens mainly through exciting a $\Delta$ resonance, followed by a $\Delta N(N)
\to NN(N)$ process (see the discussion in \cite{Lalakulich:2012cj}). It is
independent of the production process and should thus be there also in the
neutrino-induced pion production on nuclei. Astonishingly, this shape is not
seen in the pion spectra obtained by MiniBooNE
\cite{AguilarArevalo:2010bm,AguilarArevalo:2010xt}; it has been suggested that
this could be due
to a bias in the experimental analysis \cite{Katori:2012}. The high-energy
tails of the pion spectra are mainly caused by pions that were produced by DIS,
which starts at neutrino energy about 1 GeV (see fig.\ \ref{fig:pionEdep},
lower part). Mostly these high-energy  pions cascade down into
the $\Delta$ region, the remainder making up the tail \cite{Lalakulich:2012gm}.

\begin{figure}
\includegraphics[angle=-90,width=0.5\textwidth]{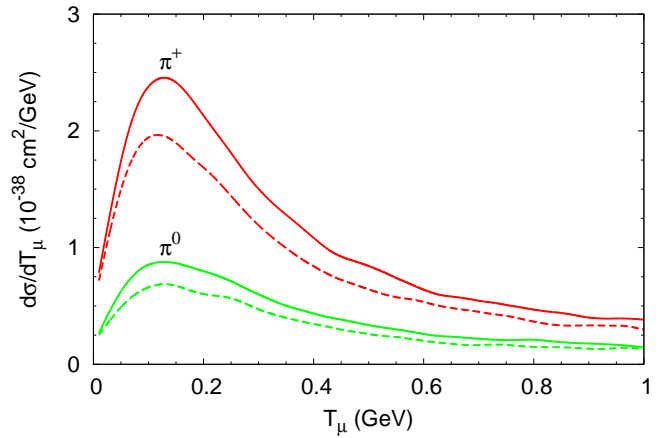}
\caption{(color online) Cross sections for $1\pi^+$ and $1\pi^0$
production as a function of the outgoing muon kinetic energy. Solid and dashed
curves are as in Fig. \ref{fig:pion-spectra}.}
\label{fig:dsdTm}
\end{figure}

In Fig.\ \ref{fig:dsdTm} we show the differential cross sections for \onepion
production as a function of the outgoing muon kinetic energy. Similar to the
results  for MiniBooNE, the cross section
peaks at a rather low energy of about 0.15 GeV with a rather flat, long tail towards higher muon kinetic energies. The peak location is roughly determined by the peak energy of the incoming neutrino beam ($\approx$ 0.6 GeV), the $\Delta$ excitation energy ($\approx$ 0.3 GeV) and its recoil energy and the muon mass ($\approx$ 0.1 GeV).

\begin{figure}
\includegraphics[angle=-90,width=0.5\textwidth]{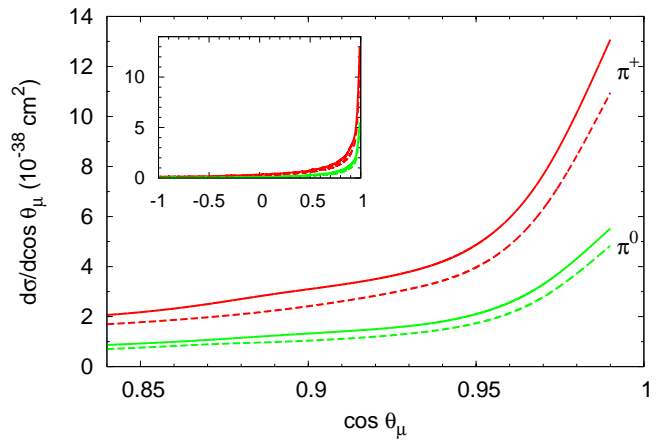}
\caption{(color online) Cross sections for $1\pi^+$ and $1\pi^0$
production as a function of the outgoing muon angle. The angular range is
restricted to that relevant for the ND280 detector. The inset shows the angular
distribution over the full angular range. Solid and dashed curves are as in
Fig. \ref{fig:pion-spectra}.}
\label{fig:dsdcm}
\end{figure}

Finally, in Fig.\ \ref{fig:dsdcm}, we show the angular distribution of the
outgoing muons with respect to the neutrino beam direction. The distribution is
strongly forward peaked. The T2K ND280 tracking detector, which is mainly
sensitive to forward angles, should thus see most of the muons.

The main uncertainty inherent in all these calculations is -- besides flux uncertainties -- the limited knowledge about the elementary pion production cross sections and, correspondingly, the $N\Delta$ axial coupling. It will, therefore, be interesting to see the experimental results from T2K. A simultaneous comparison of theory with both the MiniBooNE and the T2K data may help to disentangle the effects of flux and elementary pion production cross section.

\section{Conclusions}
Pion production in neutrino-nucleus reactions represents a major background
process to quasielastic scattering and thus influences the neutrino energy
reconstruction. As a consequence it distorts the oscillation signal. For precision studies of
oscillation parameters, such as mixing angles and mass-differences, in the long
baseline experiments it is thus important to understand this process in quite
some detail. In addition, such experiments could give useful information on the
$N\Delta$ axial coupling and form factors which are still largely unknown
\cite{Hernandez:2010bx}. Data from the T2K experiment have the potential -- when
analyzed together with the MiniBooNE data -- to answer the question of the
correct elementary cross sections for pion production. If limited to reconstructed energies up to 0.8 GeV they can give
valuable information to hadron physics. Of course, this would be even more so if
elementary targets, such as $H$ or $D$, could be employed to eliminate any
distorting effects of final
state interactions. Until such data
become available only calculations with reliable and well-tested nuclear physics based generators can be used to analyze the data.

\acknowledgements
We gratefully acknowledge a careful reading of this manuscript and useful comments by K. Mahn and H. Tanaka.

This work has been supported by DFG and BMBF.

\bibliography{nuclear}

\end{document}